\documentclass[prl,twocolumn,letterpaper, superscriptaddress]{revtex4-1}
\usepackage{graphicx}
\usepackage{dcolumn}
\usepackage{bm}
\usepackage{color}
\usepackage{tabularx}
\usepackage{array}
\usepackage{amsmath}
\usepackage{stmaryrd}  

\begin{document}

\title{Programmable Real-Time Magnon Interference in Two Remotely Coupled Magnonic Resonators}

\author{Moojune Song}
\affiliation{Materials Science Division, Argonne National Laboratory, Lemont, IL 60439, USA}
\affiliation{Department of Physics, Korea Advanced Institute of Science and Technology, Daejeon, 34141, Republic of Korea}

\author{Tomas Polakovic}
\affiliation{Physics Division, Argonne National Laboratory, Lemont, IL 60439, USA}

\author{Jinho Lim}
\affiliation{Department of Materials Science and Engineering, University of Illinois Urbana-Champaign, Urbana, IL 61820, USA}

\author{Thomas W. Cecil}
\affiliation{High Energy Physics Division, Argonne National Laboratory, Lemont, IL 60439, USA}

\author{John Pearson}
\affiliation{Materials Science Division, Argonne National Laboratory, Lemont, IL 60439, USA}

\author{Ralu Divan}
\affiliation{Center for Nanoscale Materials, Argonne National Laboratory, Argonne, IL 60439, USA}

\author{Wai-Kwong Kwok}
\affiliation{Materials Science Division, Argonne National Laboratory, Lemont, IL 60439, USA}

\author{Ulrich Welp}
\affiliation{Materials Science Division, Argonne National Laboratory, Lemont, IL 60439, USA}

\author{Axel Hoffmann}
\email{axelh@illinois.edu}
\affiliation{Department of Materials Science and Engineering, University of Illinois Urbana-Champaign, Urbana, IL 61820, USA}

\author{Kab-Jin Kim}
\email{kabjin@kaist.ac.kr}
\affiliation{Department of Physics, Korea Advanced Institute of Science and Technology, Daejeon, 34141, Republic of Korea}

\author{Valentine Novosad}
\email{novosad@anl.gov}
\affiliation{Materials Science Division, Argonne National Laboratory, Lemont, IL 60439, USA}

\author{Yi Li}
\email{yili@anl.gov}
\affiliation{Materials Science Division, Argonne National Laboratory, Lemont, IL 60439, USA}

\date{\today}

\begin{abstract}

Magnon interference is a signature of coherent magnon interactions for coherent information processing. In this work, we demonstrate programmable real-time magnon interference, with examples of nearly perfect constructive and destructive interference, between two remotely coupled yttrium iron garnet spheres mediated by a coplanar superconducting resonator. Exciting one of the coupled resonators by injecting single- and double-microwave pulse leads to the coherent energy exchange between the remote magnonic resonators and allows us to realize a programmable magnon interference that can define an arbitrary state of coupled magnon oscillation. The demonstration of time-domain coherent control of remotely coupled magnon dynamics offers new avenues for advancing coherent information processing with circuit-integrated hybrid magnonic networks.

\end{abstract}

\maketitle
In condensed matter physics, hybridization describes the formation of new eigenstates by mixing two or more excited states. This process, which has been used to describe the coupling of two atomic orbitals for chemical bonds \cite{Weinhold05}, is now being extensively cultivated in hybrid dynamic systems \cite{KurizkiPNAS2015,ClerkNPhys2020}, where dynamic excitations from disparate physical platforms are strongly coupled and form new hybrid modes in order for them to coherently exchange information while preserving the phase correlation. Hybrid dynamic systems have successfully combined diverse physical systems, such as photons , phonons, spins and magnons, and leveraged their individual advantages for implementing novel functionalities such as coherent transduction\cite{LaukQST20}, storage\cite{LuPRA13}, nonreciprocity\cite{HurstJAP22}, and sensing \cite{DegenRMP17} in quantum information science.

The use of magnetic excitations, or magnons, in coherent information processing has been well explored in magnonics\cite{ChumakNPhys2015,MahmoudJAP20,ChumakIEEE22}. Propagating magnons are controlled and engineered for spin wave signal processing, with examples of spin wave logic gate\cite{SchneiderAPL08,KlinglerAPL15,FischerAPL17}, directional coupler \cite{WangSciAdv18,WangNElectron20}, and multiplexing operations \cite{VogtNComm14,LitvinenkoarXiv2022}. These functionalities are based on the interference of two propagating magnons \cite{SchneiderAPL08,LeeJAP08,SatoAPExp2013,KlinglerAPL15,FischerAPL17,RanaPRApplied18,ChenNanoLett21}, where constructive interference leads to high amplitude or logic ``1" and destructive interference leads to low amplitude or logic ``0". However, the fundamental disadvantages of short magnon propagation distance and inefficient magnon excitations have limited the development of coherent magnon spintronics. The emerging field of hybrid magnonics \cite{LachanceQuirionAPEx2019,LiJAP2020,ZarePhysRep22,YuanPhysRep22} provides a solution to the spatial and efficiency limitations for coherent magnonic information processing. By achieving strong coupling between magnons and microwave photons \cite{SoykalPRL2010,StenningOE2013,HueblPRL2013,TabuchiPRL2014,ZhangPRL2014,GoryachevPRApplied2014,BhoiJAP2014,BaiPRL2015,LiPRL2019_magnon,HouPRL2019}, the conversion efficiency between magnons and photons can reach 100\% with cavity-enhanced magnon interactions. In addition, using microwave photons as coherent information transducer in a microwave resonator, remote coupling of two magnonic resonators can be achieved with macroscopic separation\cite{ZhangNComm2015,LambertPRA2016,BaiPRL17,YuanPRB20,LiPRL22,WangNComm22,XieQST2023}, showing the potential of building spatially distributed coherent magnonic network for interference operations.

\begin{figure}[htb]
 \centering
 \includegraphics[width=3.0 in]{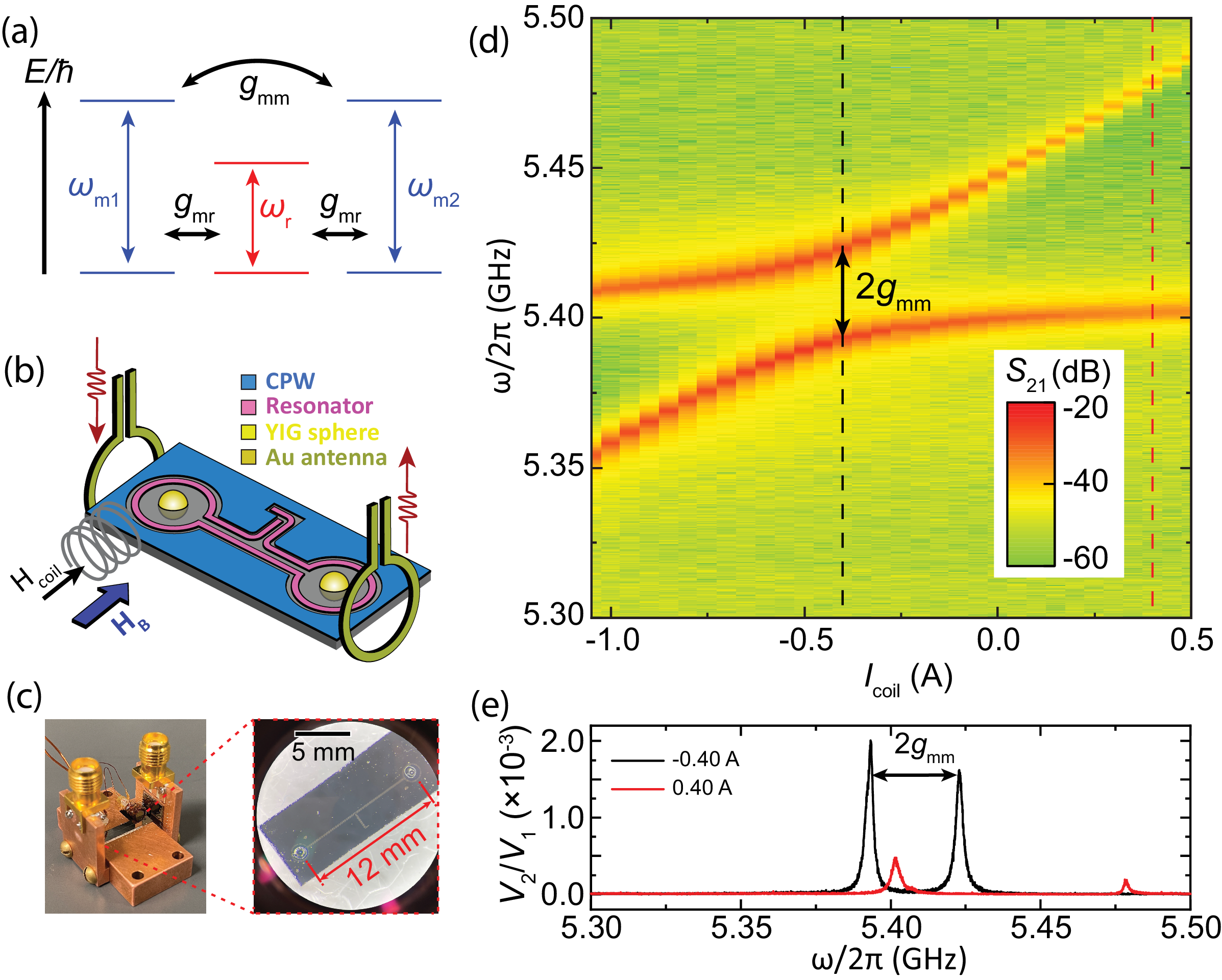}
 \caption{(a) Illustration of magnon-magnon coupling ($\omega_{m1}$, $\omega_{m2}$) via dispersive coupling to a microwave resonator ($\omega_r$). (b), Experimental schematics for the transmission measurement, with two vertical antennas adjacent to the two distant YIG spheres for selective microwave excitation and detection. (c) Pictures of the circuit setup and the superconducting resonator chip with two embedded YIG spheres ($d=250$ $\mu$m) separated by 12 mm. (d) VNA power transmission spectra from one vertical antenna to the other vertical antenna as a function of  $I_\text{coil}$. Power spectral traces for $I_\text{coil}=-0.4$ A and 0.4 A. The amplitude is plotted as $V_2/V_1 = 10^{(S_{21}/10)}$.}
 \label{fig1}
\end{figure}

In this work, we demonstrate nearly perfect constructive and destructive time-domain interference of magnon excitations between two remote yttrium iron garnet (YIG) spheres, with their strong coupling mediated by a coplanar superconducting resonator \cite{LiPRL22} (Fig. \ref{fig1}a). Using two vertical antennas (Figs. \ref{fig1}b, c), we can directly excite the magnon mode of one YIG sphere and detect the other with minimal crosstalk to the superconducting resonator. By measuring the real-time signal output from the second YIG sphere, we show that the magnon excitation from the first microwave pulse can be coherently transferred back and forth between the two YIG sphere, and interfere with the second microwave pulse excitation. The independent control of both the time and phase delay allow for arbitrary final hybrid magnonic state, ranging from constructive (2,2) or destructive (0,0) interference, to the in-phase (2,0) or out-of-phase (0,2) single mode, and to the intermediate states, thus enabling programmable control of hybrid magnonic states. Our results experimentally show that magnons preserve their full coherence by coherently transferring between remote magnonic resonators, which lays a foundation for coherent information processing with hybrid magnonics and developing functionalities in quantum magnonics.

Figure \ref{fig1}(d) shows the continuous-wave measurements of microwave transmission between the two vertical antennas ($S_{21}$) under a global field of $\mu_0 H_B=0.2$ T. A NbTi superconducting coil applies a local magnetic field to YIG sphere 1 \cite{LiPRL22} with the field direction parallel to the global field. By sweeping the coil current, clear mode anticrossing is observed between the two magnon modes, with $I_\text{coil}=-0.4$ A marking the frequency degeneracy condition for the two magnon modes ($\omega_{m1}/2\pi=\omega_{m2}/2\pi=\omega_0/2\pi=5.405$ GHz). The magnon-magnon avoided crossing yields a coupling strength of $g_{mm}/2\pi=14.8$ MHz. This is mediated by the dispersive coupling of both YIG spheres to the nearest superconducting resonator mode ($\omega_r/2\pi=5.27$ GHz) \cite{RameshtiPRB2018,GrigoryanPRB2019,XuPRB2019,RenPRB22}, with a magnon-photon coupling of of $g_{mr}/2\pi=46$ MHz. From the frequency detuning, the theoretical magnon-magnon coupling strength is \cite{LiPRL22} $g_{mr}^2/|\omega_m-\omega_r |2\pi=16.3$ MHz, which is close to $g_{mm}$ extracted from the experiments (14.8 MHz). Shown in Fig. \ref{fig1}(e), we also highlight that the microwave transmission is significantly enhanced when the two magnon modes are degenerate and strongly coupled ($I_\text{coil}=-0.4$ A), as compared with the case when they are decoupled ($I_\text{coil}=0.4$ A). This shows that additional microwave signal is transmitted from YIG sphere 1 to 2 via the superconducting resonator, as compared to the decoupled case where the transmitted signal is dominated by the free-space radiation between the two vertical antennas. The large signal-to-noise ratio of magnon excitations is crucial for observing the real-time evolution of magnon states.
\begin{figure}[htb]
 \centering
 \includegraphics[width=3.0 in]{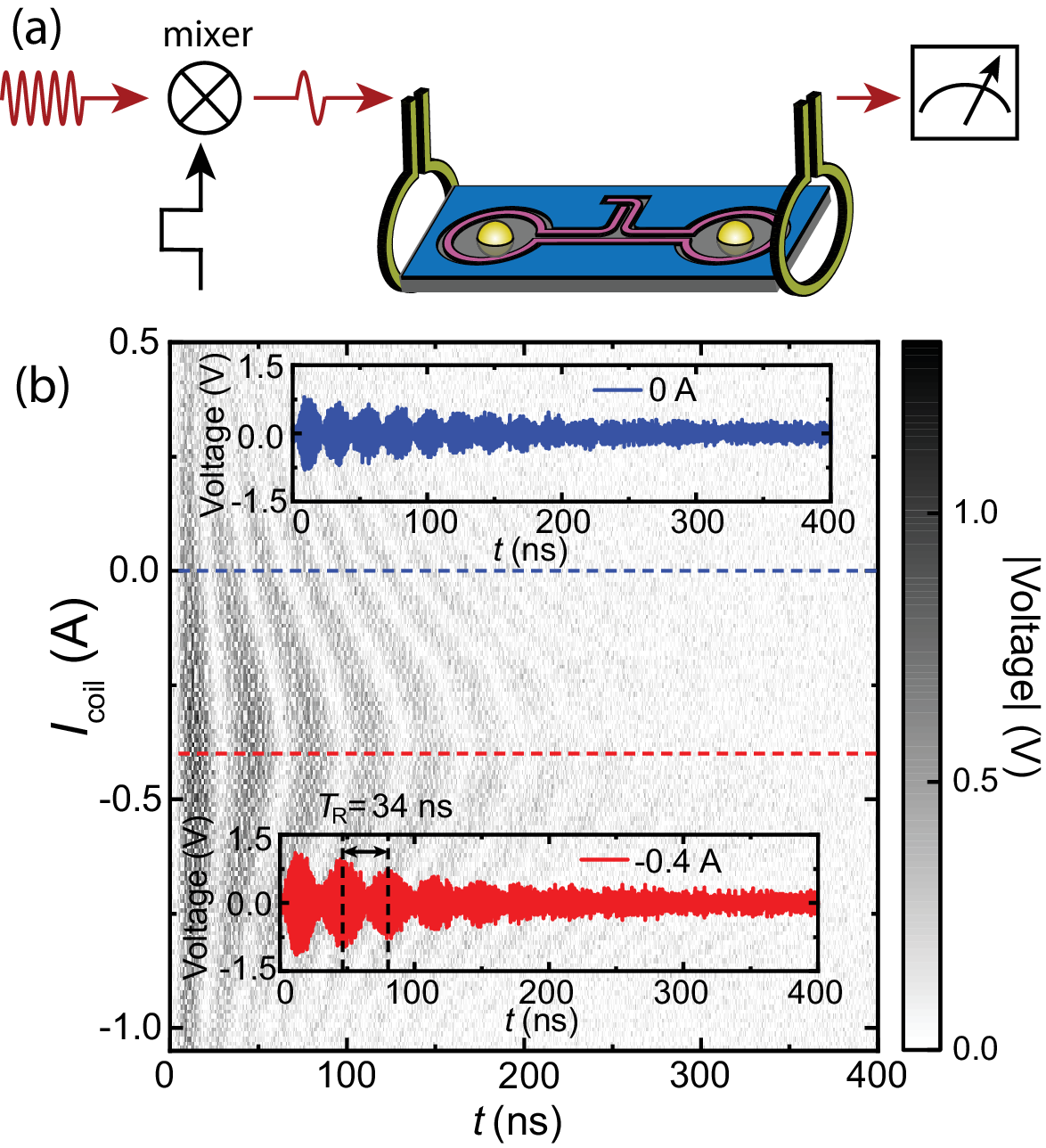}
 \caption{(a) A schematic illustration of the pulse microwave measurements, where the pulse input is generated by mixing a continuous microwave to a square function with a width of 10 ns, and the output is measured by a fast oscilloscope without any averaging (single-shot measurement). (b) Time traces of the output signals for different $I_\text{coil}$ under a global field of $\mu_0 H_B=0.2$ T. Inset: Time trace measured at $I_\text{coil}=0$ A and $I_\text{coil}=-0.4$ A, respectively.}
 \label{fig2}
\end{figure}

\begin{figure*}[htb]
 \centering
 \includegraphics[width=5.0 in]{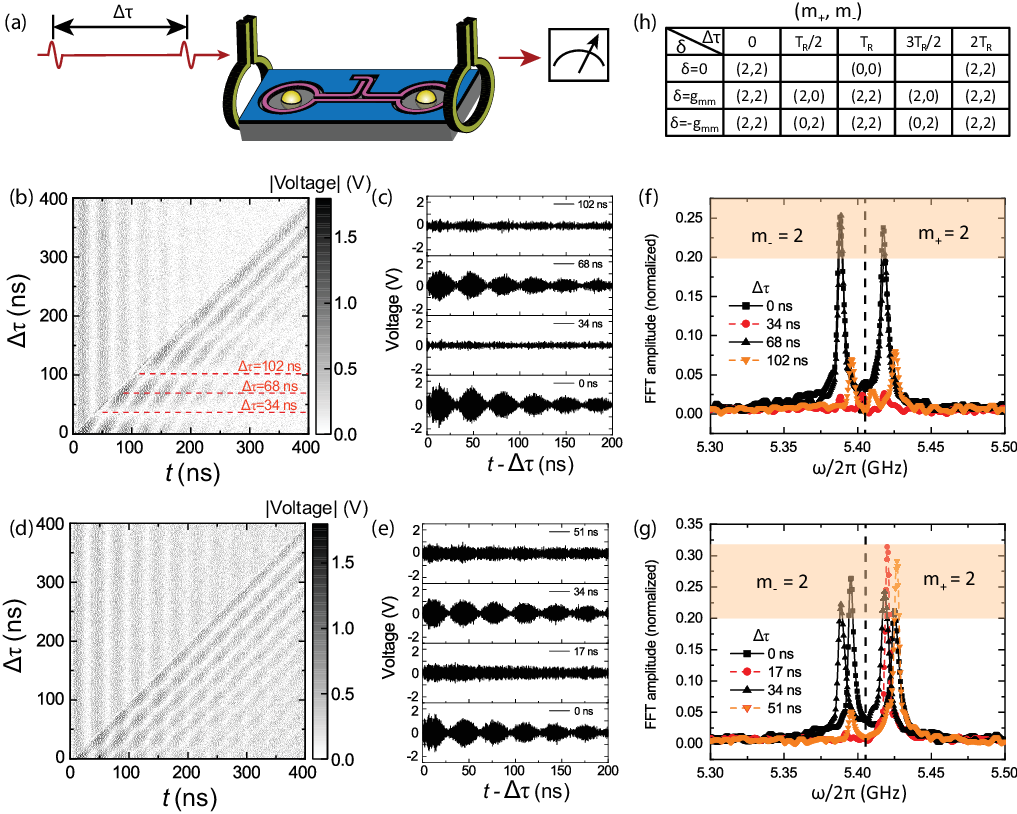}
 \caption{(a) Schematic of two-pulse excitations of coherent magnonic states. (b) Time traces of the output signals for different $\Delta\tau$ with two-pulse excitations, measured at $\omega = \omega_0$. (c) Individual time traces of (b) at $\Delta\tau=0$, 34, 68 and 102 ns, with the time $t=0$ starting right after the second pulse. (d) Same as (b) measured at $\omega = \omega_0+g_{mm}$. (e) Individual time traces of (d) at $\Delta\tau=0$, 17, 34 and 51 ns, with the time shifted by $\Delta\tau$, starting right after the second pulse. (f-g) FFT spectra of the time traces from (c) and (e), showing alternating final states (f) between (0,0) and (2,2), and (g) between (2,0) and (2,2). Orange shades indicate the range for $m_\pm=2$. Dashed vertical line at $\omega_0/2\pi=5.405$ GHz splits the regimes of $m_+$ and $m_-$. (h) Table of ($m_+$, $m_-$) states for different $\delta$ and $\Delta\tau$.}
 \label{fig3}
\end{figure*}

To investigate the real-time dynamics of the two coupled magnonic resonators, we excite the magnon mode of YIG sphere 1 with a microwave pulse at $\mu_0 H_B=0.2$ T, and measured the signal output of YIG sphere 2 by a fast oscilloscope [Fig. \ref{fig2}(a)]. The microwave pulse frequency is set to $\omega_0$ for optimal magnon excitation. The pulse duration is 10 ns and is much shorter than the magnon-magnon transduction period $\pi/g_{mm}=35$ ns, so that the pulse can be treated as an instantaneous excitation of magnons in YIG sphere 1. Shown in Fig. \ref{fig2}(b), a clear Rabi-like oscillation of magnon amplitude in YIG sphere 2 is measured at port 2. At $I_\text{coil}=-0.4$ A where the two magnon modes are degenerate [Fig. \ref{fig2}(d)], we find that the magnon population oscillates with a period of $T_R = 34$ ns, agreeing well with $\pi/g_{mm}$. The relaxation time $T_1= 161.50$ ns corresponds to a magnon damping rate of $\kappa_m/2\pi=1/(2\pi T_1)=0.98$ MHz. When the magnon frequencies are detuned, e.g. at $I_\text{coil}=0$ A, the magnon oscillation period becomes shorter [Fig. \ref{fig2}(c)] due to the larger frequency difference between the two hybrid modes, which is also reflected by the increasing magnon mode difference in Fig. \ref{fig1}(d). The quantitative analysis of the Rabi-like oscillation between the two YIG spheres is important for the study of magnon interference below.

In order to investigate magnon interference between the two remotely coupled YIG spheres, we inject two consecutive microwave pulses with the same amplitude to excite YIG sphere 1 [Fig. \ref{fig3}(a)] under the condition of $I_\text{coil}=-0.4$ A and $\mu_0 H_B=0.2$ T ($\omega_{m1}=\omega_{m2}=\omega_0$). Because the microwave pulses are generated by mixing a continuous-wave microwave signal with square waves, they maintain strict phase coherence, which provides a good phase reference for studying coherent magnon interactions. In the first example, we set the pulse microwave frequency to be equal to the magnon frequency as $\omega=\omega_0$. Fig. \ref{fig3}(b) shows the evolution of two-pulse interference as a function of time for different $\Delta \tau$. The diagonal boundary defines where the excitation of the second pulse starts. Before the second pulse, the magnon excitation in YIG sphere 2 shows Rabi-like oscillation defined by $T_R$, which is the same as measured in Fig. \ref{fig2}(b). However, after the second pulse, an interference pattern emerged, showing near-perfect construction or destruction of the Rabi-like oscillation. In particular, the time trace shows maximal amplitude which is twice the amplitude of single-pulse excitation at $\Delta\tau=2nT_R$, and near-zero amplitude at $\Delta\tau=(2n+1)T_R$ [Fig. \ref{fig3}(c)]. This shows that the magnon excitation maintains its coherence after being fully transduced back and forth between two spatially separated YIG spheres, and can interfere with the second microwave pulse with its full amplitude. By slightly changing the frequency of the microwave pulses from $\omega_0$ to $\omega_0+g_{mm}$, as shown in Fig. \ref{fig3}(d), the interference pattern has completed changed. The period of the interference pattern is reduced in half, with strong and weak magnon outputs at $\Delta\tau=2n(T_R/2)$ and $\Delta\tau=(2n+1)(T_R/2)$, respectively [Fig. \ref{fig3}(e)]. The strong output still shows Rabi-like oscillation with twice the maximal amplitude as the single-pulse excitation. However, for the weak output, the Rabi-like oscillation disappears and the magnon excitation shows monotonous relaxation with time, with an exponential relaxation time of  $T_1^*= 161.50$ ns, same as $T_1$ measured above. Same feature is observed in the case of $\omega=\omega_0-g_{mm}$; see the Supplemental Information for more details \cite{supplement}.

The magnon interference can be analytically described by the time evolution of two coupled magnon resonators $\vec{m_1}$ and $\vec{m_2}$. Their hybridized eigenmodes can be formulated as:
\begin{equation}\label{eq02}
  \vec{m_\pm}(t) = {1\over \sqrt{2}}(\vec{m_1}+\vec{m_2})e^{-i(\omega_0\pm g_{mm})t}
\end{equation}
where $\vec{m_+}$ and $\vec{m_-}$ denote the in-phase and out-of-phase modes, respectively, with eigenvalues of $\omega_0\pm g_{mm}$. The magnon state can be labelled by the amplitudes of the two eigenmodes, ($m_+$, $m_-$). After the first microwave pulse, which excite only the dynamics in YIG sphere 1, the magnon state evolves from the initial state (0,0) to (1,1), where the amplitudes of the two eigenmodes have the same weight and are renormalized as one. The change of states happens after the second pulse. By controlling $\Delta\tau$ and the frequency detuning $\delta = \omega-\omega_0$, the final magnon state can be modified to arbitrary combination with the amplitude range between 0 and 2 as a result of destructive or constructive interference. Here $\delta$ introduces a phase shift between the microwave pulse and the magnon dynamics as $\Delta\phi = \delta\Delta\tau$. The final state after two pulses can be derived as:
\begin{equation}\label{eq03}
  m_\pm(\Delta\tau) =  \left|1+e^{-i(\delta\mp g_{mm})\Delta\tau}\right| = 2\left|\cos{(\delta\mp g_{mm})\Delta\tau \over 2}\right|
\end{equation}
The table in Fig. \ref{fig3}(h) lists the final magnon states for a few different $\delta$ and $\Delta\tau$. In the case of $\delta=0$ (Fig. 3b), Eq. (3) is reduced to $m_\pm (\Delta\tau)= 2|\cos (g_{mm} \Delta\tau/2)|$, which leads to synchronized constructive (2,2) or destructive (0,0) interference at $g_{mm} \Delta\tau=2n\pi$ or $g_{mm} \Delta\tau=(2n+1)\pi$. In the case of $\delta=g_{mm}$, Eq. (3) is reduced to $m_+=2$, $m_- (\Delta\tau)=2|\cos (g_{mm} \Delta\tau)|$, leading to a constant amplitude of $m_+$ and an oscillating $m_-$ with half the oscillation period. The theoretical predictions of Eq. (\ref{eq03}) nicely agree the experiments in Figs. \ref{fig3}(b-e), with a Rabi frequency of $\pi/g_{mm}=33.8$ ns that matches with $T_R=34$ ns. We also conduct the fast-Fourier transform (FFT) of the time traces from  Figs. \ref{fig3} (c,e) and show the peaks in Figs. \ref{fig3} (f,g), respectively. The peaks are located at $\omega_0\pm g_{mm}$, indicating the magnon states of $m_+$ and $m_-$. At different $\Delta\tau$, the magnon states oscillates between (0,0) and (2,2) for $\delta=0$ and between (2,0) and (2,2) for $\delta=g_{mm}$, which are indicated by the similar FFT peak height around 0.25 a.u. (orange shade, $m_\pm=2$) or zero peak (($m_\pm=0$)) at $\omega_0\pm g_{mm}$. We also note that weak peaks appear when the magnon state is supposed to be zero at longer delay time, e.g. $\Delta\tau=102$ ns in Fig. \ref{fig3}(f) and $\Delta\tau=51$ ns in Fig. \ref{fig3}(g). This is due to the magnon relaxation and decoherence with $\Delta\tau$, as will be discussed later. Similarly, when $\delta=-g_{mm}$, according to Eq. (\ref{eq03}), the magnon states oscillate between (0,2) and (2,2); see the Supplemental Materials for details \cite{supplement}.


\begin{figure}[htb]
 \centering
 \includegraphics[width=3.0 in]{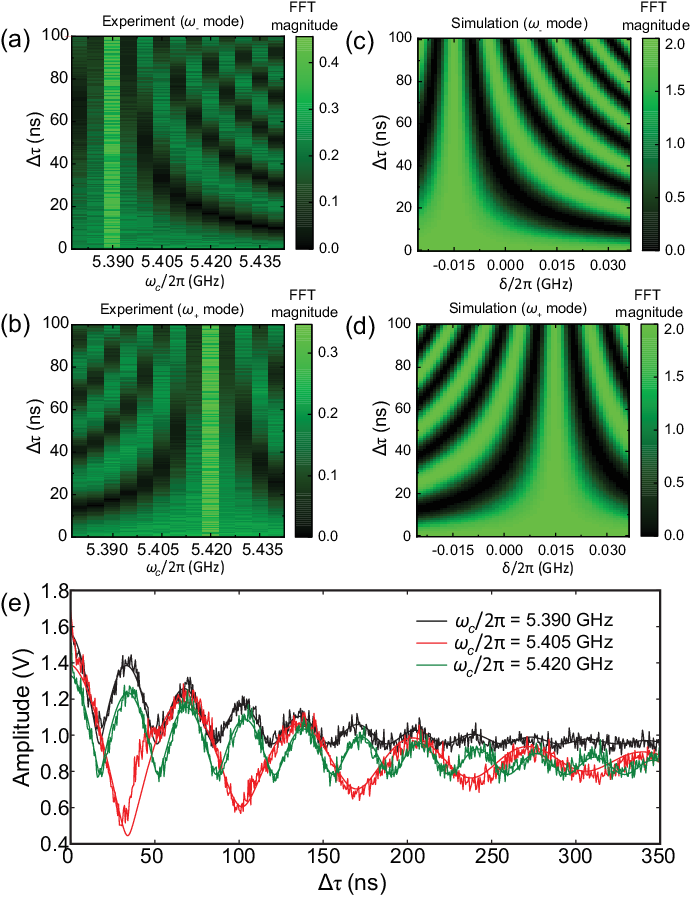}
 \caption{(a-b) Magnon states (a) $m_+$ and (b) $m_-$ extracted from the FFT of the time traces measured at different $\omega_c$ and $\Delta\tau$. Colorbars denote the FFT peak amplitude with examples shown in Fig. \ref{fig2}(e) and (f). (c-d) Theoretical predictions of (c) $m_+$ and (d) $m_-$ from Eq. (\ref{eq03}) using $g_{mm}/2\pi=14.3$ MHz and $\omega_0/2\pi=5.405$ GHz. (e) Plot of $V_{max}$ as a function of $\Delta\tau$ for $\omega_c/2\pi=5.390$, 5.405 and 5.420 GHz. The fit equation is $V_{max} = V_0[m_+(\Delta\tau)+m_-(\Delta\tau)]$ by taking $V_0$, $g_{mm}$ and $T_2$ as fit parameters.}
 \label{fig4}
\end{figure}

To further demonstrate the capability of programming magnon states with two-pulse interference, we show in Figs. \ref{fig4}(a) and (b) the experimentally extracted final state of $m_+$ and $m_-$ as a function of $\omega$ and $\Delta\tau$. The value of each pixel corresponds to the FFT amplitude of the time trace after two microwave pulses. Clear interference patterns are observed and nicely match with the calculation of Eq. (\ref{eq03}) as shown in Figs. \ref{fig4}(c) and (d). As a characteristic of the mode selection, for $\omega/2\pi=(\omega_0-g_{mm})/2\pi=5.390$ GHz [$=(\omega_0+g_{mm})/2\pi=5.420$ GHz], $m_+$ ($m_-$) always shows the maximum value of 2 while $m_-$ ($m_+$) oscillates between 0 and 2, as also shown in Figs. \ref{fig3}(e) and (g). It can be derived that any magnon state ($m_+$, $m_-$) between (0,0) and (2,2) can be achieved for $0\leq \Delta\tau \leq \pi/g_{mm}$ and $0\leq \delta \leq g_{mm}$; see the Supplemental Materials for details \cite{supplement}.

The magnon interference also allows us to quantify the magnon decoherence time $T_2$ during the remote magnon transduction process. Different from $T_1$ which characterizes the relaxation time of magnon states, $T_2$ describes how long magnons preserve their phase correlation for interference. The two-pulse experiments set the condition for magnon-magnon interference with a delay time $\Delta\tau$ for free magnon evolution. The external microwave source, which is used to create the two consecutive pulses, also defines an excellent phase reference for magnon interference analysis. In addition, the hybrid magnonic system enables a dephasing test of nonlocal magnon excitations, which is directly related to coherent magnon information processing. In Fig. \ref{fig4}(e), we plot the maximal output voltage amplitude after the second pulse, $V_{max}$, as a function of $\Delta\tau$ for $\delta=0$ (red) and $\delta=\pm g_{mm}$ (black and green). $V_{max}$ represents $m_++m_-$. The dephasing process can be described by adding the dephasing term in Eq. (\ref{eq03}):
\begin{equation}\label{eq04}
  m_\pm(\Delta\tau) = \left|1+e^{-i(\delta\mp g_{mm})\Delta\tau-\Delta\tau/T_2}\right|
\end{equation}
The fitting curves to Eq. (\ref{eq04}) nicely reproduce the experimental results for all the three frequency detunings. The extracted $T_2=139.0$ ns is slightly smaller than $T_1=161.5$ ns, suggesting that the magnon spin dynamics are highly phase coherent due to their exchange coupling. We note that since the magnon-magnon interference is conducted in the dispersive magnon-photon coupling regime, neither $T_1$ nor $T_2$ are sensitive to the resonator damping rate. The two-pulse interference process is similar to Ramsey interference of a single spin with two $\pi/2$ pulses \cite{GaetaPRL16}. For $\delta=0$, the spin-up and spin-down states are represented by (0,0) and (2,2) states, with one pulse setting the magnon state to (1,1) and two pulses driving the magnon states to oscillate between (0,0) and (2,2) depending on $\Delta\tau$. This oscillating state can be restricted to $m+$ or $m_-$ only for $\delta=- g_{mm}$ or $g_{mm}$, respectively. We note another work which also measures magnon Ramsey interference \cite{XuPRL21}. The difference is that our work directly characterizes the interference of pure magnon hybrid states whereas in Ref. \cite{XuPRL21} the dephasing process occurs with magnon-photon hybrid states. 

In conclusion, we have demonstrated real-time interference of two remotely coupled magnon resonators (YIG spheres) that are embedded and dispersively coupled to a coplanar superconducting resonator. By controlling the frequency and time delay of two microwave pulses, we can obtain arbitrary combination of the hybrid magnon states ($m_+$, $m_-$) from constructive or destructive magnon interference excited by the two pulses. From the time-delay dependence of magnon interference, we also obtain the magnon dephasing time which is close to the magnon relaxation time. Our results provide a realistic circuit-integrated hybrid magnonic system for implementing coherent magnon operations in the time domain. The control of hybrid magnon states can be used to realize remote magnon-magnon entanglement and distributed magnon quantum gate operations by integrating superconducting qubits on the same superconducting circuit \cite{LachanceQuirionScience2020,WolskiRPL20,XuDaPRL23}. Additionally, the two-magnon-resonator system can be extended towards coupled magnon networks with distributed magnon oscillators, which is applicable to many magnonic computing ideas with ultra-long magnon coherence time, tunable magnon frequency and coupling strength.

\textit{Acknowledgments.} We thank Wolfgang Pfaff, Juliang Li, Volodymyr G. Yefremenko and Margarita Lisovenko for discussion and support on the experiment. This work was supported by the U.S. DOE, Office of Science, Basic Energy Sciences, Materials Sciences and Engineering Division under contract No. DE-SC0022060. Work performed at the Center for Nanoscale Materials, a U.S. Department of Energy Office of Science User Facility, was supported by the U.S. DOE, Office of Basic Energy Sciences, under Contract No. DE-AC02-06CH11357. K.-J.K. is supported by KAIST-funded Global Singularity Research Program for 2021 and the National Research Foundation of Korea (NRF) funded by the Korean Government (MSIP) under grant No. 2020R1A2C4001789, 2016R1A5A1008184. M.S. was supported by the education and training program of the Quantum Information Research Support Center, funded through the National research foundation of Korea (NRF) by the Ministry of Science and ICT (MSIT) of the Korean government under grant No. 2021M3H3A103657313.


%

\end{document}